\documentclass[aip,% jmp,
 amsmath,amssymb,
preprint,
Conference Proceedings
]{revtex4-1}

\usepackage{graphicx}% Include figure files
\usepackage{dcolumn}% Align table columns on decimal point
\usepackage{bm}% bold math
\usepackage{textcomp}
\usepackage{verbatim}

\usepackage{xr}

\newcommand{\beginsupplement}{%
        \setcounter{table}{0}
        \renewcommand{\thetable}{S\arabic{table}}%
        \setcounter{figure}{0}
        \renewcommand{\thefigure}{S\arabic{figure}}%
        \setcounter{section}{0}
        \renewcommand{\thesection}{S\arabic{section}}
     }
\usepackage[utf8]{inputenc}
\usepackage[T1]{fontenc}
\usepackage{mathptmx}
\usepackage{gensymb}

\draft % marks overfull lines with a black rule on the right

\begin{document}
%To get a quick word count:

%\quickwordcount{Edits}
%To get a quick character count:
%\quickcharcount{Main}
%To get a detailed word count:
%\detailtexcount{Edits}

\title{Simultaneous three-axis torque measurements of micromagnetism} %Title of paper



\author{K.R. Fast}

\author{J.A. Thibault}
\affiliation{University of Alberta, Department of Physics}

\author{V.T.K. Sauer}
\affiliation{University of Alberta, Department of Physics}
% \affiliation{Wyvern Space, Edmonton}

\author{M.G. Dunsmore}

\author{A. Kav}
\affiliation{University of Alberta, Department of Physics}

\author{J.E. Losby}
\affiliation{University of Alberta, Department of Physics}
\affiliation{University of Calgary, Department of Physics and Astronomy}
\affiliation{Nanotechnology Research Centre (NANO), National Research Council Canada (NRC) }

\author{Z.Diao}
\affiliation{University of Alberta, Department of Physics}
\affiliation{Florida Agricultural and Mechanical University, Department of Physics}

\author{E.J. Luber}
\affiliation{University of Alberta, Department of Chemistry}

\author{M. Belov}
\affiliation{Nanotechnology Research Centre (NANO), National Research Council Canada (NRC) }

\author{M.R. Freeman}
\affiliation{University of Alberta, Department of Physics}

%\thanks{}
%\affiliation{University of Alberta}


\date{\today}

% Have just copied the abstract that was submitted for the 
% conference, we'll probably want to shorten it for the manuscript. 
\begin{abstract}
Measurements of magnetic torque are most commonly preformed about a single axis or component of torque. Such measurements are very useful for hysteresis measurements of thin film structures in particular, where high shape anisotropy yields a near-proportionality of in-plane magnetic moment and the magnetic torque along the perpendicular in-plane axis. A technique to measure the full magnetic torque vector (three orthogonal torque components) on micro- and nano-scale magnetic materials is introduced. The method is demonstrated using a modified, single-paddle silicon-on-insulator resonant torque sensor. The mechanical compliances to all three orthogonal torque components are maximized by clamping the sensor at a single point. Mechanically-resonant AC torques are driven by an RF field containing a frequency component for each fundamental torsional mode of the device, and the resulting displacements read out through optical position-sensitive detection. The measurements are compared with micromagnetic simulations of the mechanical torque to augment the interpretation of the signals.  As an application example, simultaneous observations of hysteresis in the net magnetization along with the field-dependent in-plane anisotropy is highly beneficial for studies of exchange bias.
    
% More generally, any component of cross-product torque involves a linear combination of perpendicular moment and magnetic susceptibility, yielding the rotational magnetic anisotropy or stiffness of the spin texture in response to torque. \par % Additional experimental inputs are required when the goal is to disentangle moments and susceptibilities. \par 

% Figure \ref{fig:SEM} shows a completed device supporting a  cobalt oxide / permalloy bilayer disk, deposited on a silicon paddle which is supported by a single beam.  
%Deposition and patterning of the magnetic material is performed with the mechanical device doubly-clamped.  In the final processing step, the torsion mechanical susceptibilities to $x$- and $z$-torques are increased substantially by releasing one of the clamping points with a small focussed ion beam cut.

% A typical hysteresis measurement obtained as an in-plane DC bias field is swept is shown in Figure \ref{fig:hysteresis}.\par 

\end{abstract}


\maketitle

\section{Introduction}
\label{sect:Intro}

%Intro needs piles of work. 
Torque magnetometers\cite{Beck1918} have returned as a popular tool in micromagnetism over recent decades, developing from micromechanical cantilevers for atomic force microscope tips\cite{Albrecht1998}. These cantilevers were adopted as the basis of torque magnetometers \cite{Rossel1996, Heydon1997}, owing to their high sensitivity to low-mass magnetic materials and their ability to measure magnetic anisotropies \cite{Brugger1999}. \par 

One advantage of torque magnetometers is their versatility. The readout of mechanical deflection can be accomplished in a number of ways, the most common of which being capacitive and piezoresistive \cite{Brugger1999, Kohout2007, Rossel1998}. Techniques have been developed with these readout methods to measure torque about two orthogonal axes of the cantilevers. These techniques have advanced through maximizing the sensitivity along both axes \cite{Kohout2007}, measurement of force on nanomechanical resonators \cite{Rossi2017}, as well as development of sensitivity to all three orthogonal torque axes \cite{Hoon2000} although to-date, only two orthogonal torque modes have been measured concurrently. Torque measurements are most commonly performed on in-plane modes. Measurements of the out-of-plane torque, as accomplished by Hajisalem et al. \cite{Hajisalem2019}, allow for the study of in-plane magnetic anisotropy of thin-film structures. Pairing the out-of-plane torque with both in-plane modes will enable more detailed characterization of magnetic samples, as seen in other forms of magnetometry (e.g. MOKE microscopy \cite{Mehrnia2019}). We introduce here a method of concurrent measurement of the full magnetic torque vector. \par

A micromechanical torque magnetometer consists of a magnetic specimen mounted on a flexible support structure, typically a cantilever. The magnetometer measures the deflection of the support due to a torque applied on the magnetic sample through the interaction of its moment with an external field,
\begin{equation}
    \mathbf{\tau} = \mathbf{m}\times \mu_0 \mathbf{H}.
\end{equation}
Torque magnetometry is capable of measuring DC or AC torques. The technique presented here studies AC magnetic torques, achieved through the application of an alternating field perpendicular to a DC magnetizing field. When the applied field is driven at a frequency corresponding to the support's mechanical resonance, the sensitivity of the sensor to applied torque is maximized.  \par 

% Measurement of the complete torque vector allows for detailed characterization of a magnetic sample. Utilizing this technique with additional experimental inputs will permit future experimental determination of the magnetic susceptibility tensor. 
To measure the torque about three orthogonal axes, a device susceptible to resonance about each axis is required as a supporting structure. Such a device and the technique used to simultaneously extract each torque component is presented, with emphasis on interpretation of results attained in such a fashion.

\section{Experimental Details}
\label{sect:Experimental}
To demonstrate our method of measuring the complete torque vector, we use single-paddle, silicon-on-insulator devices, as shown in Figure \ref{fig:SEM}. Details of the fabrication process are given in the Supplementary Materials \ref{supp:sect:Fabrication}. Each torque mode supported by the device corresponds to a unique mechanical resonance of the sensor.  A simple modification of a standard device recipe\cite{Diao2013} makes the sensitivity to $x$- and $z$- torques comparable to that for $y$-torque. 
%This permitted mechanical susceptibility to torque about the $x$- and $z$- axes, as defined by the coordinate system of Figure \ref{fig:SEM}, in addition to the $y$-axis torque available in the original device configuration.  Figure \ref{fig:SEM} shows an SEM image of a modified device, with arrows indicating the mechanical motion about each axis. 
 % The modification of these devices occurred after investigations of the flexural torque mode \cite{Losby2015,FaniSani2014}.\par % These resonances are in the radio frequency (RF) range, with each mode well-decoupled from the others. \par 
\begin{figure}
    \begin{center}
    \includegraphics[scale=0.68]{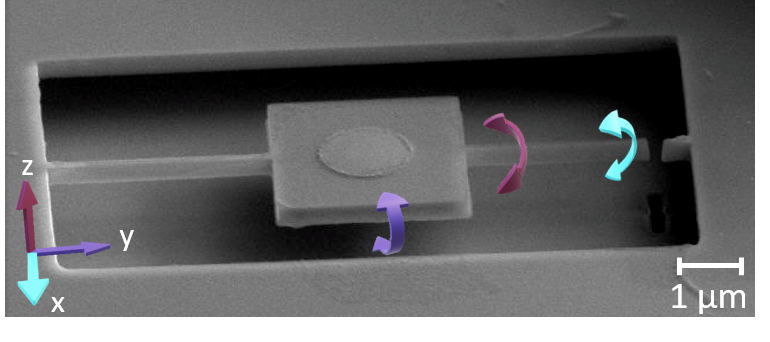}
    \caption{Electron micrograph of a micromechanical device used for demonstration of the three-axis torque measurement technique.  The arrows indicate the three orthogonal torsional displacements. The light blue arrow overlaps the position of a cut made to release the paddle and allow susceptibility about three orthogonal axes. 
    \label{fig:SEM}}
    \end{center}
\end{figure}
An RF field is applied to the sensor by driving a wound coil with a current containing components of each of the sensor's resonant frequencies. This coil is placed asymmetrically atop the device to produce in-plane field components at the sample in addition to the primary out-of-plane field produced by the coil. The positioning of the coil produces RF field primarily along the $x$- and $z$- axes. A torque is applied to the sample by the application of a DC magnetizing field in conjunction with the applied RF field. This DC field is produced by a permanent neodymium iron boron magnet. Hysteresis measurements of torque are collected via translation of this magnet along the system's $x$-axis.\par 
%\textit{The strength of the three RF field components at the sensor has yet to be determined. This determination, once made, will allow for a more complete description of the system.}
The signal readout is accomplished with a free-space optical interferometer using a helium neon laser ($\lambda$ = 633 nm). The three frequency components in the interferometer signal are separated with a Zurich Instruments multi-channel lock-in amplifier, permitting measurement of signal magnitude and phase for each torque mode. The amplitude of the recorded signal is converted to physical units through thermomechanical calibration, following the procedure described by Hauer et al. \cite{Hauer2013}. Thermomechanical calibration data and the corresponding torque sensitivities are presented in the Supplementary Materials \ref{supp:sect:TM}. %The procedure to calibrate to thermal motion of the sensor involves using the quality factor of the thermally activated resonance and the effective mass for a specific resonant frequency to determine the conversion of the signal to mechanical units. \par 

The use of optical interferometry to read out torque signal additionally allows a method with which to characterize the mechanical motion of the sensors. This is done through scanning the laser spot across an area encompassing the device to measure the reflectance, signal strength, and signal phase. More details on mapping the signal motion through interferometry is given in the Supplementary Materials \ref{supp:sect:Raster}.

\section{Results} 
\label{sect:Results}

\subsection{Hysteresis}
\label{sect:Hysteresis}
\begin{figure}
    \begin{center}
    \includegraphics[scale=0.5]{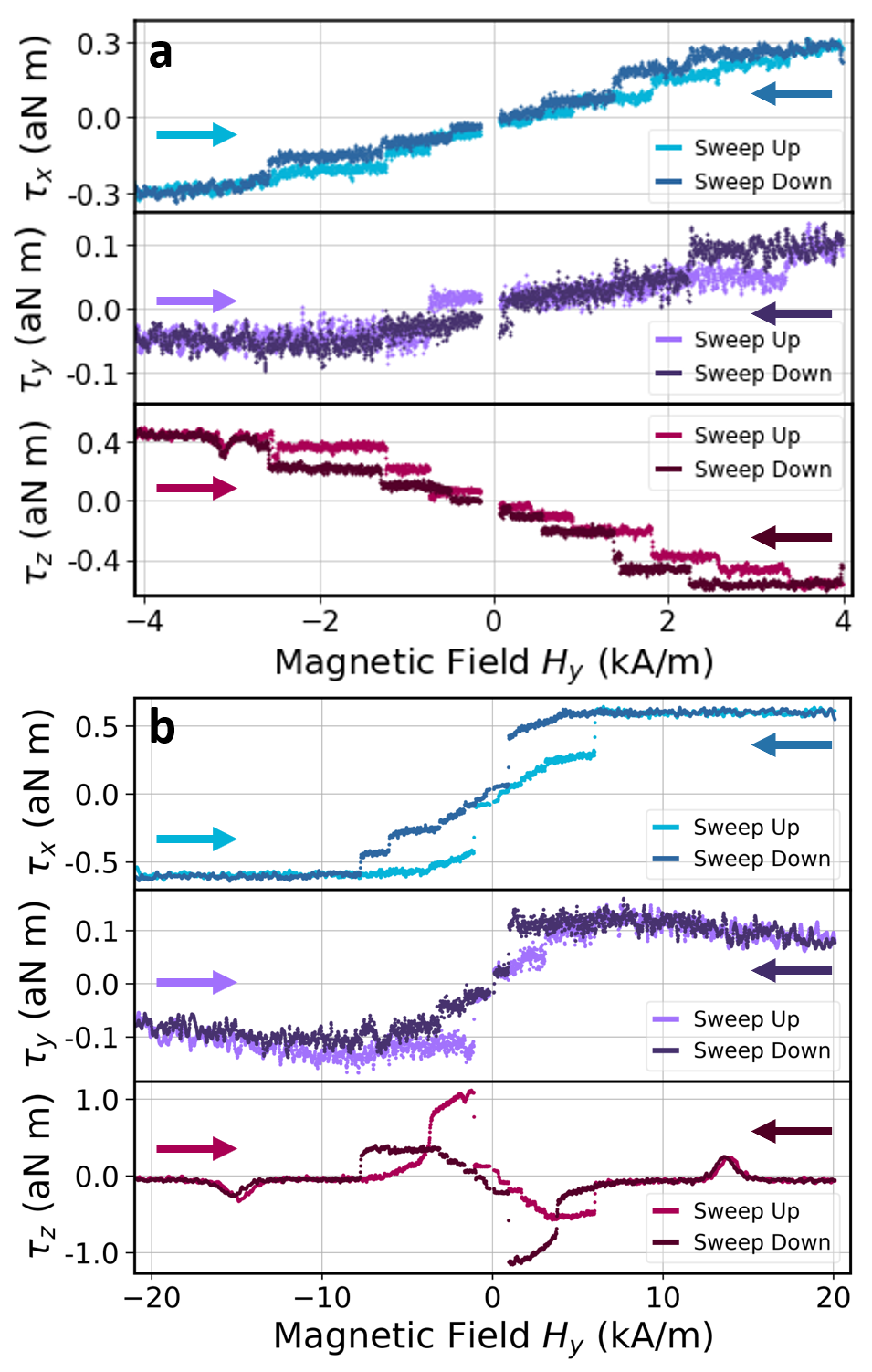}
    \caption{Simultaneous hysteresis measurements of orthogonal AC torques from a cobalt oxide / permalloy bilayer disk at room temperature at (top to bottom) $f_{mechanical}$ = 1.886, 4.194, 0.953 MHz.  The in-plane bias field direction here is rotated $104.4^{\degree}$ from $\hat{x}$ towards $\hat{y}$. Solid arrows indicate the direction of changing field for each branch of the hysteresis loop. (a) A minor hysteresis loop, where the magnetization in the permalloy layer remains in a vortex configuration. (b) A major hysteresis loop, representing the behaviour of the torque as the permalloy layer adjusts between the saturated and vortex magnetization states. The magnitude of torque was determined through thermomechanical calibration \cite{Losby2012}. \label{fig:hysteresis}}
    \end{center}
\end{figure}

Measurements to illustrate the effectiveness of our three-axis measurement technique are shown in Figure \ref{fig:hysteresis}, where torque about the three orthogonal axes of our sensor was measured simultaneously. A typical hysteresis measurement on this apparatus is taken with the DC field oriented along the $x$-axis of the coordinate system defined in Figure \ref{fig:SEM}. Such a configuration maximizes the $y$-torque and minimizes the $z$-torque to the point where it is indistinguishable from noise. This is due in large part to the small magnitude of in-plane RF field produced by the off-axis coil, as well as a negligible in-plane shape anisotropy throughout the permalloy disk. An out-of-plane magnetic torque relies on the contributions of in-plane field and magnetization. These are produced by the off-axis RF coil, as well as a small out-of-plane DC field caused by an asymmetry between the height of the magnet and sample. These contributions are typically of similar magnitude and dwarfed by out-of-plane torque contributions in $\tau_x$ and $\tau_y$. Cancellation of the two in-plane contributions from the cross product typically negates the out-of-plane torque. It has been experimentally determined that in low field, the cancellation of the contribution terms is minimized for a DC field rotated 104.4$\degree$ from the positive $x$-axis. At this angle, the magnetization state in low-fields produces a strong $z$-torque contribution.  \par 
The data shown in Figure \ref{fig:hysteresis} was collected in ambient conditions with the laser spot located at the position noted by a red circle in Figure \ref{fig:raster}a and the permanent magnet rotated 104.4$\degree$ from the positive $x$-axis.% The magnitudes of each torque mode were determined through thermomechanical calibration, as described in Supplementary Materials \ref{supp:sect:TM}.  \par 

% , due to the small magnitude of in-plane RF field produced by the off-axis coil, as well as cancellation of the two in-plane torque contributions \cite{Hajisalem2019}. The data in Figure \ref{fig:hysteresis} was collected with the permanent magnet rotated 104.4 degrees counterclockwise from the positive x-axis, producing a negative DC field primarily along the y-axis. This field configuration resulted in a maximized z-torque in low field, allowing for detailed characterization of the magnetic specimen. \par 
The low-field behaviour, presented as a minor hysteresis loop in Figure \ref{fig:hysteresis}a, is indicative of the Barkhausen effect, which has been thoroughly described in previous works \cite{FaniSani2014,Burgess2013,Burgess2014}. The Barkhausen features of Figure \ref{fig:hysteresis}a are also clearly seen in Figure \ref{fig:hysteresis}b between -1 and +5 kA/m (Sweep Up) and between +1 and -5 kA/m (Sweep Down). The characteristics of this low-field state are indicative of a vortex magnetization state, where the magnetization in the permalloy disk is circularly oriented in-plane around an out-of-plane magnetized vortex core. As the field applied to the sample is adjusted, the vortex core translates along the disk, interacting with defects on the permalloy layer. This interaction results in the pinning of the vortex core within the defect. As such, the magnetization of the disk gets trapped in this pinning site, resulting in the plateaus seen throughout Figure \ref{fig:hysteresis}a.  Notably, the slopes of the plateaus in the $x$- and $z$-torques of Figure \ref{fig:hysteresis} do not agree. The $x$-torque plateaus possess a small slope, while those of the $z$-torque remain flat. This difference in behaviour is indicative of a deformable vortex with a pinned core, as described by Burgess et al. in their deformable vortex pinning model \cite{Burgess2014}. 

% these plateaus in the torque display behaviour of a rigid vortex. This is evident in the lack of slope within these plateaus, where a sloped plateau is indicative of a flexible vortex state, as described by Burgess et al. in their deformable vortex pinning model \cite{Burgess2014}.

% would be present in a disk with a flexible vortex \cite{Burgess2014}.% \textit{The slope of these pinning sites is important and something I need to discuss, but I need to figure out exactly what's going on there before I can do this because I don't fully understand the history / importance of the slope???}\par 

The behaviour of the torques throughout the major hysteresis loop of Figure \ref{fig:hysteresis}b provides key insight into the change in magnetization throughout the loop. This loop was initialized in negative field at a 104.4$\degree$ magnet angle. The behaviour of the $x$- and $z$-torques in this major loop are particularly descriptive of the system's magnetization. At -4 kA/m as the field is swept up, the slope of the $x$-torque abruptly changes. This slope is constant until -1 kA/m, where the magnetization jumps significantly into a new magnetization state (the low-field vortex state depicted in Figure \ref{fig:hysteresis}a). From this state, the vortex annihilates, resulting in a saturated magnetization. A similar trend is seen on the sweep-down in field. The $z$-torque shows similarly distinct magnetization changes at the same transition fields. At the first transition, the $z$-torque abruptly changes from a zero to non-zero (approximately 1 aN m) magnitude, which linearly increases until the magnetization drops into the vortex state where Barkhausen effects are evident. When the vortex state is annihilated at 6 kA/m, the $z$-torque is once again negligible. These distinct changes in magnetization and corresponding torque behaviour can be characterized and described through application of micromagnetic simulation. The small bumps in $z$-torque seen at $\pm$ 14 kA/m are a result of thermally-assisted dynamics of small closure domains in the spin texture. Magnetic edge roughness results in pinning sites, between which these closure domains jump, giving rise to a shift in signal phase. The phase corresponding to the $z$-torque reveals 180 degree peaks corresponding to these features, indicating that these thermal dynamics are at play. These peaks are found to have a strongly field-angle dependent fingerprint, consistent with edge roughness. 

% Pinning sites caused by magnetic edge roughness permit hopping of these closure domains between

% Hopping between pinning sites caused by magnetic edge roughness. This effect gives rise to a shift in signal phase. Investigation of the $z$-torque signal phase reveals a 180 degree phase-shift corresponding to these features, indicating the thermal dynamics at play. These peaks ar

% activation in the permalloy spins. This is verified through investigation of the signal phase, recorded with the lock-in amplifier. The phase shows 180 degree peaks at the same fields, which is a strong indicator of a thermally activated signal. Through further investigation of the system, these peaks are seen to be highly angle-dependent, with the precise location of the peaks shifting with the angle of the applied DC field. 
%
% These changes in the data are caused by a change in the magnetization state throughout the loop. The specific magnetization states corresponding to these changes will be discussed with reference to simulations in Section \ref{sect:Sim}. 

\subsection{Simulation}
\label{sect:Sim}

Micromagnetic simulations using Mumax \cite{Vansteenkiste2014} are a vitally useful tool to investigate the magnetization of the sample throughout the hysteresis loops of Figure \ref{fig:hysteresis}. Torques are calculated in post-processing to directly compare simulations with data. To correctly interpret the mechanisms behind specific features in the data, simulations which isolate specific physical parameters are applied. Through this method, the key features of the minor hysteresis loop of Figure \ref{fig:hysteresis}a and the major hysteresis loop of Figure \ref{fig:hysteresis}b can be isolated. The simulations discussed here employed an applied field rotated 104.4$\degree$ to mimic the DC field applied throughout the loops of Figure \ref{fig:hysteresis} with no edge smoothing applied to investigate the effects of edge roughness. The magnetic area was defined a permalloy cylinder with 20 nm depth and 1.44 $\mu$m diameter. At ambient temperatures, the cobalt oxide is assumed to contribute negligibly and is as such neglected. At each DC field value (stepped in increments of 0.08 kA/m), the field was dithered along the (111) direction to mimic an RF driving field. The amplitude of the driving field was assumed to be 0.02 kA/m in post-processing of the torques. This magnitude was chosen to match the $x$-torque simulation and data magnitudes, under the assumption that the calibrated RF field magnitude should agree with simulation values.  \par 
\begin{figure}
    \centering
    \includegraphics[scale=0.48]{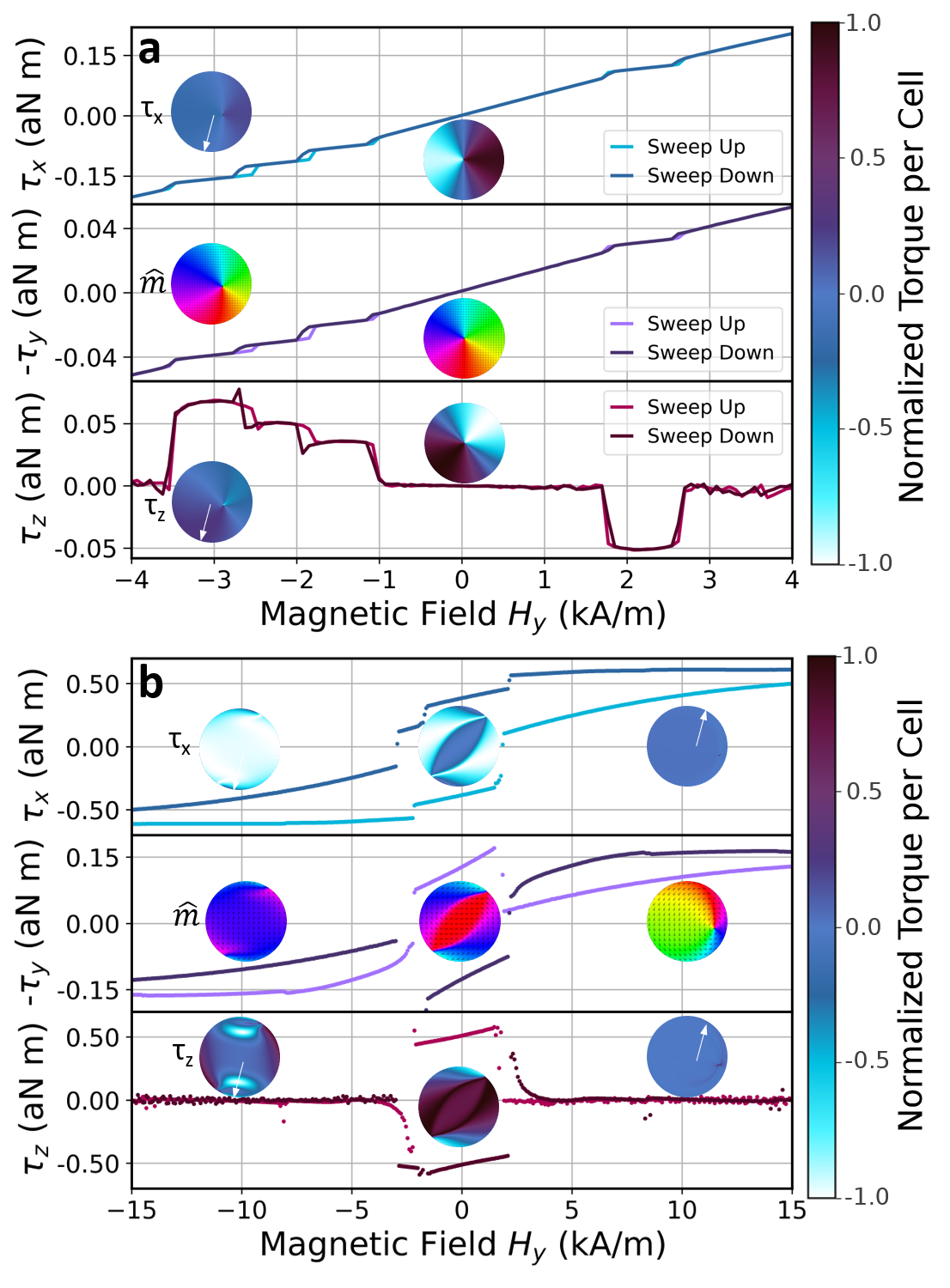}
    \caption{Simulations to explain behaviours seen in Figure 3. a) The simulation is constrained to a vortex state, with artificial defects implanted along the vortex core's path. b) A simulation to probe the torque and magnetization states by extending the fields to achieve saturation in the permalloy magnetization. The circular insets in the $\tau_x$ and $\tau_z$ sectors represent spatial maps of the torque at a) -3, 0 kA/m b) -10, 0, +10 kA/m. The white arrow in the torque spatial maps indicates the direction of the DC field. The insets in the $\tau_y$ sector show spatial maps of the magnetization at the same fields. The colour wheel associated with the magnetization spatial maps is the standard CMYK colour wheel.   }
    \label{fig:simulation}
\end{figure}

For the simulation presented in Figure \ref{fig:simulation}a, a vortex was initialized, and the field swept over a small range. The simulation used 2.8 nm cells over a 512 pixel wide grid. Four artificial defects were implanted in the permalloy cylinder along the path of the vortex core to produce pinning sites. These defects were created as small square regions (20 nm wide) with a saturation magnetization 85\% of the permalloy saturation magnetization. The artificially implanted defects produced in-plane torque behaviour ($\tau_x$ and $\tau_y$) corresponding to Barkhausen jumps, as seen in Figure \ref{fig:hysteresis} and as described in previous works \cite{FaniSani2014,Burgess2013}. The out-of-plane torque behaviour shows a dependence on defect position for the torque amplitude. Evidence of hysteresis due to pinning is found in all three torque modes. However, this hysteresis only appears for jumps between neighbouring pinning sites, and does not affect the behaviour as the vortex core jumps from a pristine platform into a defect, or vice versa. \par 

The corresponding normalized spatial maps of the $x$- and $z$-torque components supplement the simulations of Figure \ref{fig:simulation}. The normalized torque magnitude in each cell is indicated by the colour bar on the right hand side of the figure. White arrows in the maps indicate the direction of the DC field. The $y$-torque mode sectors contain spatial maps of the net magnetization in the permalloy disk, utilizing the standard CMYK colour wheel. The torque maps at zero field in Figure \ref{fig:simulation}a indicate the expected integration to zero torque. Similarly, the torque maps at -3 kA/m (corresponding to a pinned vortex core) cannot be integrated to zero. The out-of-plane torque at the vortex core is significantly larger than that of the unpinned core. Correspondingly, the net out-of-plane torque in the pinning site has a significant non-zero value, as expected. \par 

These torque and magnetization maps are helpful in visualizing the magnetization state throughout the major and minor hysteresis loops, particularly with respect to the more complex magnetization states seen in Figure \ref{fig:hysteresis}b. The simulation in Figure \ref{fig:simulation}b involves a pure permalloy disk (i.e. no artificial defects) on a 256 cell wide grid with 5.6 nm cells, initialized in a negatively saturated magnetization state. The corresponding magnetization and torque maps reveal a "transition" magnetization state between the saturation and vortex states. This transition state takes the form of a "cat's eye", with two cores located on opposite corners of the disk, between which an oblong region of constant magnetization occurs. The magnetization stays in this configuration over a narrow field range before transitioning into the vortex state. This state develops as the magnetization gets trapped on rough pixellated edges of the disk, disallowing the sample from evolving directly into the vortex state from saturation. This transition state causes hysteresis in the vortex state as a result of the difference in magnetization prior to vortex nucleation. This hysteresis appears as an "opening" of the low-field vortex configuration between the sweep-up and sweep-down in field. This opening is in contrast with the expected vortex behaviour exhibited in Figures \ref{fig:hysteresis}a and \ref{fig:simulation}a, where no opening is observed, and is in agreement with behaviour near zero-field in Figure \ref{fig:hysteresis}b. \par 

The features corresponding to each magnetization state in Figure \ref{fig:simulation}b  share characteristics with the features noted in Figure \ref{fig:hysteresis}b. In particular, the slopes in torque during the transition magnetization state appear in stark contrast to the saturation behaviour of the sample. The vortex nucleation and annihilation fields are not expected to align between simulation and measurement, due in large part to the absence of thermal contributions in simulation. The similarities noted strongly indicate the existence of the cat's-eye magnetization state between -4 and -1 kA/m in Figure \ref{fig:hysteresis}b. 

%\textit{Torque map procedure:} The circular insets of Figure \ref{fig:simulation} are mapped images of the simulated torque at specific DC field magnitudes (-125, 0, and +125 G). The torque is calculated at 5 fields during the dither process throughout the simulation, calculated as $\tau_k = m_iB_j - m_jB_i$, and saved as an ovf file. The ovf file is converted to a csv for analysis. The torque map at each saved point in the dither field is loaded into a 5xNxN array for an NxN simulation grid. The torques are multiplied by Msat*V$_{cell}$ for the volume-averaged torque of each cell. The slope of torque vs field for each grid index is calculated, and saved in a new NxN array. This array multiplied by the approximated drive amplitude (1G) is plotted as an image to create the torque map.

%additional comment to add after the discussion of the behaviours in the data. 
%The comparison between the dominant in-plane torque alongside the out-of-plane torque enables confirmation of the magnetic behaviour. The $z$-torque behaviour, co-analyzed with the $y$-torque yields clear interpretations, whereas the $z$-torque co-analyzed with the $x$-torque alone would be inconclusive. 

%some words about why this is important/what information this gives us. 
\section{Conclusions} 
\label{sect:Conclusions}
Measurement of three orthogonal torque axes provides a particularly useful tool to describe the properties of magnetic samples. Comparison with micromagnetic simulation provides a distinct platform with which to confidently describe these properties and their effect on hysteresis loops of the sample. \par 

The technique to simultaneously measure three orthogonal torque components will be put to use as a tool to further investigate behaviours of these sensors. Particularly, the sensor will be studied at low temperatures. The ferromagnetic / antiferromagnetic interface between the permalloy and cobalt oxide layers have exhibited exchange bias at low temperatures. Work to investigate this effect is underway, utilizing the measurement of the complete torque vector. Comparison between the behaviours of the sample at the ambient conditions presented here and those at low temperatures will be of particular use in the investigation of exchange bias. %In this investigation, micromagnetic simulations will continue to be utilized as a method to determine physical properties present in the sample. 

\section*{Supplementary Material}
See Supplementary Material for sample fabrication details, scanned signal maps, and thermomechanical calibration data.

% % If you have acknowledgments, this puts in the proper section head.
\begin{acknowledgments}
The authors gratefully acknowledge support from the Natural Sciences and Engineering Research Council of Canada (RGPIN 04239), the Canada Foundation for Innovation (34028), and the Canada Research Chairs (230377). The nanomechanical torque devices were created using fabrication tools of the University of Alberta nanoFAB and the National Research Council Nanotechnology Research Centre.  
\end{acknowledgments}

\section*{Data Availability}
The data that support the findings of this study are available from the corresponding author
upon reasonable request.
% % Create the reference section using BibTeX:
% \newpage
\bibliography{bib.bib}

\newpage
\beginsupplement
\section*{Supplementary Materials}
\section{Sample Fabrication Details}
\label{supp:sect:Fabrication}

These devices were fabricated at the University of Alberta nanoFAB facility as a double-clamped paddle supporting a thin cobalt oxide / permalloy bilayer disk. A subsequent modification of the devices involved the release of one supporting arm with a focused ion beam (FIB) cut.

Depositions were done using confocal magnetron sputtering on an ATC Orion 8 sputter system (AJA International Inc.), using 2" diameter targets (Plasmaterials Inc.). Prior to deposition, the base pressure of the chamber was less than 2.0E-7 Torr. The CoO films (20 nm thick) were reactively RF sputtered from a Co target at a power of 118 W, a sputtering gas pressure of 4 mTorr, flow rate of 14 SSCM Ar and 6 SCCM O2 and a deposition rate of 0.228 nm/min. After reactive sputtering the CoO the chamber was filled with argon to a pressure of 40 mTorr and held for 60 seconds, then pumped back down to vacuum for 60 seconds. This process was repeated 5 times to ensure that any residual oxygen from the reactive sputtering had been evacuated from the chamber prior to NiFe deposition. The NiFe film (20 nm thick) was co-deposited from Ni and Fe targets and DC power supplies at 40 and 13 W, with deposition rates of 1.13 and 0.31 nm/min respectively and a sputtering pressure of 4 mTorr using argon gas flowing at 20 SCCM. Deposition rates very determined using a quartz crystal monitor, where the CoO was assumed to have a density of 6.44 g/cm$^3$ and the bulk densities of 8.91 and 7.86 g/cm3 for Ni and Fe were used.

\section{Scanned Signal Maps}
\label{supp:sect:Raster}

\begin{figure}
    \centering
    \includegraphics[scale=0.4]{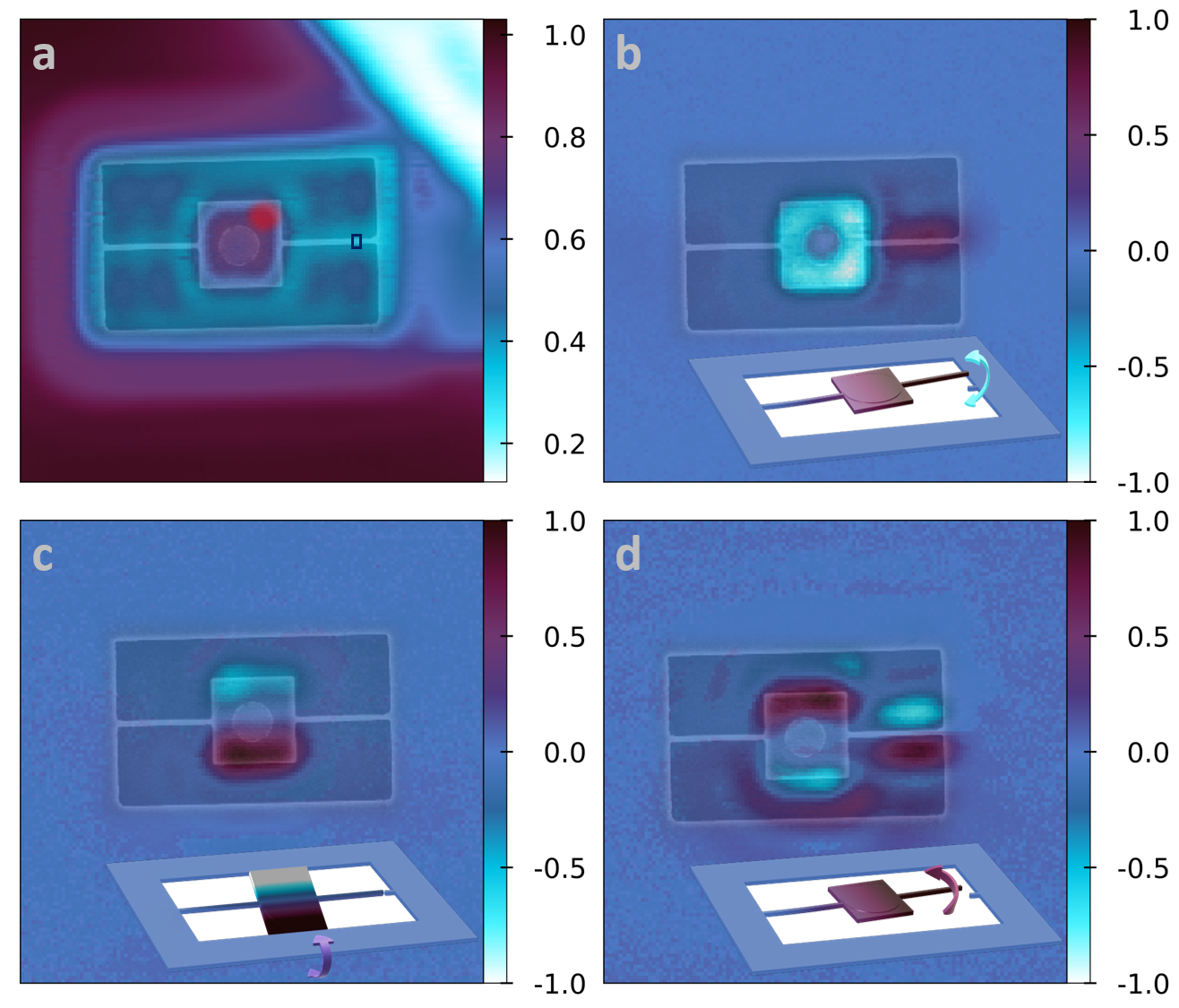}
    \caption{Scanned images of the normalized a) reflected intensity of optical light, b) x-torque ($f_{mechanical}$ = 1.886 MHz), c) y-torque ($f_{mechanical}$ = 4.194 MHz), d) z-torque ($f_{mechanical}$ = 0.953 MHz), overlaid with a transparency of an SEM image of the sample. The dark square on the right supporting arm of the SEM in (a) indicates the approximate location of the FIB-cut, which was performed after the collection of this SEM. The red circle on the top right of the paddle in (a)  indicates the approximate laser spot position for the collection of the data shown in Figure \ref{fig:hysteresis}. Simulated motion corresponding to each resonant frequency is shown at the bottom of (b), (c), and (d), with arrows indicating the motion of each mechanical mode.}
    \label{fig:raster}
\end{figure}

The optical reflectance and the motion of the device when driven at each mechanical frequency are shown through signal maps in Figure \ref{fig:raster}. Each quadrant includes a transparent overlay of an SEM image of the device as a reference for the sensor orientation. This SEM was taken prior to the FIB cut. A dark rectangle on the arm in Figure \ref{fig:raster}a indicates the location of the cut. The red circle on Figure \ref{fig:raster}a indicates the laser spot position for the data collection throughout the main text. Finite element simulation of each torque mode performed in COMSOL Multiphysics is shown at the bottom of each panel as a reference for the expected motion of each torque mode.  

The scanned reflectance in Figure \ref{fig:raster}a produces a detailed image of the device. The top right corner of this panel shows a remnant of photoresist from the fabrication procedure. The low reflectance in this region is likely an effect of this remnant sitting at non-normal angle relative to the silicon, resulting in the reflection of light away from the photodetector. The frequency-dependent signals of Figure \ref{fig:raster}b-d show no evidence of this remnant material and indicate motion consistent with expectation. \par 

The reflectance and signal phase of the $x$- and $z$-torque modes in Figure \ref{fig:raster} show notable features as effects of diffraction. The reflectance from the arms is less obvious than the features of the paddle, and the phase behaviours of the $x$- and $z$-torques (Figure \ref{fig:raster}b and \ref{fig:raster}d) are not indicated by finite-element simulations. These features stem from diffracton effects related to the laser beam width (approximately 0.8 $\mu$m) in comparison with the arm width (0.2 $\mu$m). The $y$-torque (Figure \ref{fig:raster}c) scan does not show evidence of diffraction effects, due largely to the negligible motion of the arms. These diffraction effects are seen around the paddle (3.0 x 2.9 $\mu$m) edges, though are not prominent in the paddle's center, when the beam spot is positioned completely on the paddle. \par 
Simulation of the $x$-torque indicates a uniform phase across the device. The $x$-torque shown in Figure \ref{fig:raster}b shows an unexpected phase difference between the paddle and the arm. This phase reversal is a result of scattered reflected light from the surface below the device, which eliminates reflected intensity from out-of-plane motion of the cut arm.  Similarly, the motion of the $z$-torque in Figure \ref{fig:raster}d exhibits a 180 degree phase difference between the paddle and the arm. In this case, diffraction losses from the edges of the arm, where the laser spot overlaps the arm and the undercut are the cause of this phase difference. The diffraction losses surpass the magnitude of the arm's reflected intensity, lending to an optical readout of negative motion from in-plane modulation of the sensor. 

\section{Thermomechanical Calibration}
\label{supp:sect:TM}
\begin{figure}
    \centering
    \includegraphics[scale=0.5]{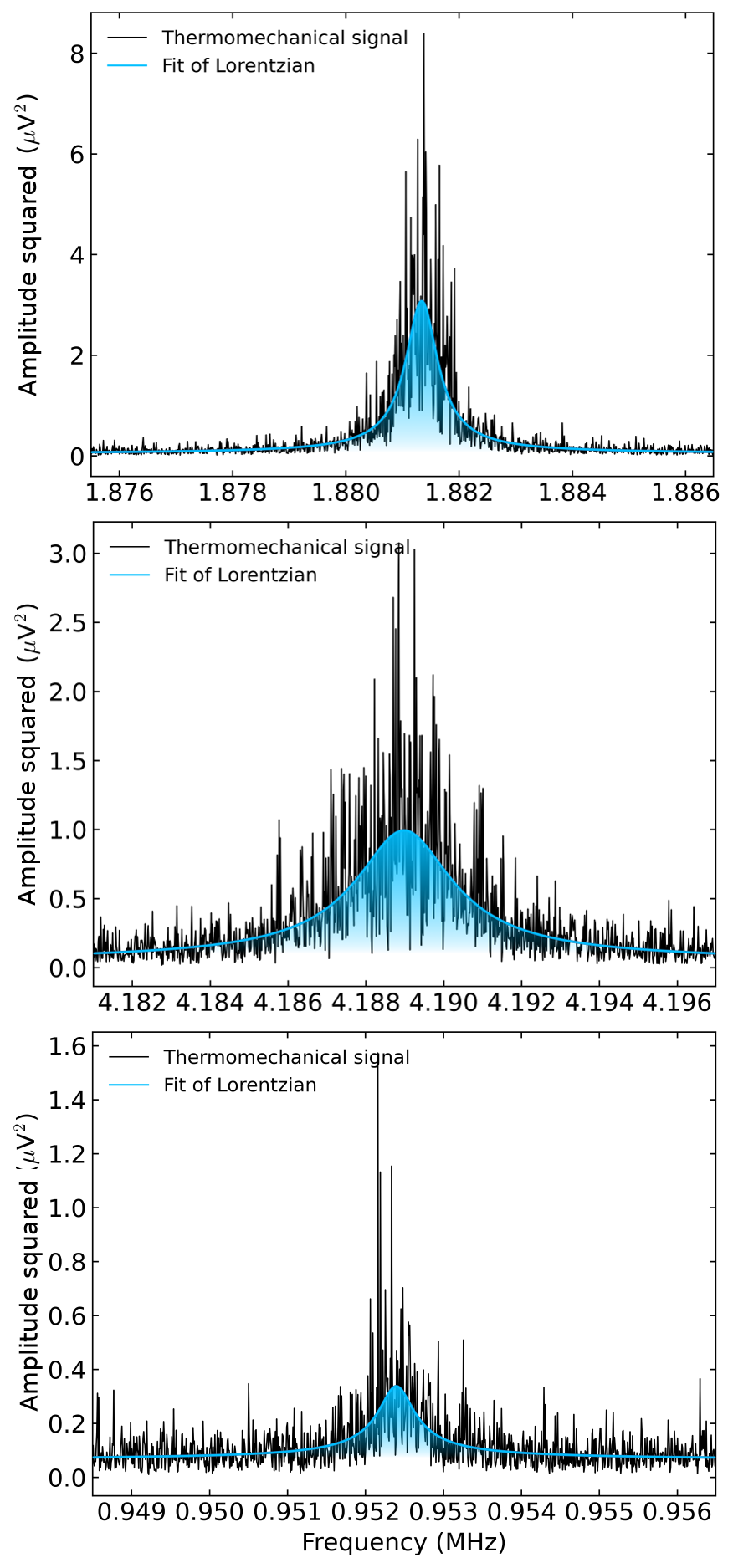}
    \caption{Thermomechanical signal for each torque mode, with a Lorentzian fit applied. The parameters of the Lorentzian fit were used to determine the sensitivity, as per the procedure by Hauer et al. \cite{Hauer2013}.}
    \label{supp:fig:TM}
\end{figure}

The sensitivities to each torque mode were determined through thermomechanical calibration by measuring torque with no applied RF field. The thermomechanical signal for each mode is given in Figure \ref{supp:fig:TM}. The quality factor of the resonances is paired with material parameters and sensor dimensions to calculate the sensitivity as per the procedure of Hauer et al. \cite{Hauer2013}. These sensitivities are calculated to be 16.50, 67.72, 48.23 aN m / mV for $\tau_x$, $\tau_y$, and $\tau_z$, respectively.
%The procedure to calibrate to thermal motion of the sensor involves using the quality factor of the thermally activated resonance and the effective mass for a specific resonant frequency to determine the conversion of the signal to mechanical units. \par 

\end{document}